\documentclass[12pt]{article}
\usepackage{amsfonts}
\usepackage{latexsym}
\usepackage{amsmath}
\usepackage{amssymb}
\usepackage{longtable}
\usepackage{epsfig}

\newcommand{\eps}{\epsilon}
\newcommand{\beps}{\bar{\epsilon}}
\newcommand{\tlambda}{\tilde{\lambda}}

\begin{document}

\begin{titlepage}

\begin{flushright}
  UB-ECM-PF-07-12
\end{flushright}

\begin{center}
\vspace{3cm} \baselineskip=16pt {\LARGE \bf Non-propagating degrees
of freedom in supergravity and very extended $G_2$}
 \vskip 1 cm
{\large  Joaquim Gomis and Diederik Roest} \\
\vskip 10 mm {\small
    Departament Estructura i Constituents de la Materia \\
    Facultat de F\'{i}sica, Universitat de Barcelona \\
    Diagonal 647, 08028 Barcelona, Spain \\
   E-mail: {\tt \symbol{`\{}gomis, droest\symbol{`\}}@ecm.ub.es}}
\end{center}

\bigskip
\centerline{ABSTRACT}
\bigskip\bigskip

Recently a correspondence between non-propagating degrees of freedom in maximal supergravity and the very extended algebra $E_{11}$ has been found. We perform a similar analysis for a supergravity theory with eight supercharges and very extended $G_2$. In particular, in the context of $d=5$ minimal supergravity, we study whether supersymmetry can be realised on higher-rank tensors with no propagating degrees of freedom. We find that in this case the very extended algebra fails to capture these possibilities.

\end{titlepage}

\section{Introduction}

The interplay between supergravities and their associated Kac-Moody algebras has received a
great amount of attention over the years.

An important first step was the discovery of hidden symmetries
\cite{Cremmer-Julia, Julia81} upon reduction to lower dimensions. In
three dimensions, one obtains gravity coupled to a scalar coset $G /
H$. Further reduction to two dimensions leads to a symmetry which is
the affine extension $G^+$, analogous to the Geroch
group $SL(2,\mathbb{R})^+$
  for pure gravity \cite{Julia82, Geroch, Breitenlohner:1986um}.
 In one dimension the relevant symmetry is expected to be the over
 extension $G^{++}$ \cite{Julia82, Nicolai}.
 The latter has mainly been considered in the context of eleven-dimensional supergravity near space-like singularities and $E_{10} = E_8^{++}$
 \cite{damourhenneauxnicolai, damournicolai}, see \cite{tenforms1} for IIB. In this framework, space-time is expected
 to arise from the dynamics of a $\sigma$-model in one dimension. In addition, there is a conceptually
 different approach based on the non-linear realisation of (the conformal group together with)
 the very extension\footnote{The simultaneous non-linear realization
 of the affine group and the conformal group in four dimensions reproduces the Einstein equation
 of general relativity \cite{Borisov}.} $G^{+++}$, like $E_{11} = E_8^{+++}$ for the $d=11$ theory
 \cite{West1, West2} as well as the IIB theory \cite{West3}.

Very recently, the relation between the non-propagating degrees of
freedom of supergravity, closure of the supersymmetry algebra and
the corresponding Kac-Moody algebras has come into focus. In
particular, in \cite{gaugings1, gaugings2} it was shown
how all the mass deformations and possible gaugings of maximal supergravity in $d \geq 3$ dimensions\footnote{It would be interesting to see if the recent results on gaugings in $d=2$ of \cite{Samtleben} can be incorporated in $E_{11}$ as well.}, or rather the $(d-1)$-forms dual to these constants, correspond to specific
generators in the very extended algebra $E_{11}$. An exception must be made here for gaugings that violate the action principle, as will also be discussed in section \ref{discussion}. In addition,
$E_{11}$ makes predictions for the possible multiplets for the
$d$-forms on which the superalgebra can be realised \cite{tenforms1,
tenforms2}. Although these forms do not carry any propagating
degrees of freedom, they are part of the field content of the theory
and play a crucial role in the story of space-time filling branes
\cite{Halbersma, Riccioni}. The possible $d$-forms that are allowed by the superalgebra have been
explicitly calculated in the cases of IIB \cite{IIB-superalgebra}
and IIA \cite{IIA-superalgebra} and found to agree with the $E_{11}$
predictions.

The  philosophy underlying the recent papers \cite{gaugings1,
gaugings2} can be summarised as follows. Given any very
extended algebra, one can decompose its adjoint representation into
representations of a Lie subalgebra $SL(d)$ (the 'gravity line').
These are labelled by their level $l$ in the Kac-Moody algebra. Up
to some level $l$, the resulting generators are interpreted as the
$d$-dimensional space-time fields of the corresponding supergravity.
Generators at higher level are interpreted as space-time fields with
more than $d$ indices, and these may correspond to dual formulations
of lower-level fields or non-propagating degrees of freedom
\cite{dual}. Given this dictionary, one can read off the possible
$(d-1)$- and $d$-forms for any very extended algebra and compare this to the closure
of the supersymmetry algebra on such forms. This extends the results of \cite{very extended}
for propagating degrees of freedom to the non-propagating $(d-1)$- and $d$-forms.

The previous ideas have also been applied to less than very
extended algebras. For example, the propagating degrees of freedom of supergravity theories can likewise be obtained
from the affinely extended $G^+$, see e.g.~\cite{Keurentjes} for a
detailed account. In addition, the overextended algebras $G^{++}$
can contain generators corresponding to the $(d-1)$-forms. An
example in $d=10$ for the overextended $E_{10}$ can be found in
\cite{Kleinschmidt-Nicolai}. However, only the very extended
$G^{+++}$ may capture all non-propagating degrees of freedom. Roughly speaking, the less than very extended
 algebras seem to be 'too small' to contain both $(d-1)$- and $d$-forms.

An interesting question is whether it is possible to extend the
striking results for $E_{11}$ to cases based on other very extended algebras.
In other words, do other very extended algebras also
predict the correct $(d-1)$- and $d$-forms for the associated
supergravity theory in $d$ dimensions? This will necessarily be in
the context of supergravities with less than maximal supersymmetry
(as maximal supergravities are associated to $E_{11}$), while the
very extended algebras will be based on other Lie algebras $G$ than
the most exceptional $E_8$. Hence all other cases are far less restricted by symmetries.
An obvious and worthwhile question is whether the correspondence found for $E_{11}$ also holds for these
less symmetric situations and if not, what the requirements are for it to hold or
what the reasons of its failure are.

In this note we address this question in the context of minimal $N=2$ 
pure supergravity in $d=5$. This theory is similar to
$d=11$ supergravity in a number of respects, see
e.g.~\cite{Mizoguchi1}: for instance, its bosonic field content only
contains a metric and a $(d-2)/3$-form $A$ with Chern-Simons term $A
\wedge dA \wedge dA$. Clear differences are that it has only 8
instead of 32 supercharges, and it reduces to the coset $G_2 /
SO(4)$ in three dimensions, see e.g.~\cite{Cremmer99}. Hence the
relevant very extended algebra is $G_2^{+++}$ instead of $E_{11}$.
It is of interest to see whether this affects the correspondence
between the non-propagating degrees of freedom and the very extended
algebra. To this end we first consider the supersymmetry algebra of
this theory and see on which $(d-1)$ and $d$-forms this can be
realised. It turns out that the allowed $(d-1)$-forms transform as a
triplet under the $SU(2)$ R-symmetry. Afterwards we compare this
with the predictions from very extended $G_2$ and finish with a
discussion of our results.

\section{Minimal supergravity in $d=5$}

We use the conventions of \cite{ungauged, gauged}.
Our metric is mostly plus. Curved (flat) indices are denoted by
Greek (Latin) letters $\mu, \nu\, \ldots$ ($m,n,\ldots$). The index
$i = 1,2$ labels the two symplectic anti-commuting fermions and is
raised and lowered according to $\psi_\mu^i = \varepsilon^{ij}
\psi_{\mu j}$ and $\psi_{\mu j} = \psi_{\mu}^i \varepsilon_{ij}$
with $\varepsilon_{12} = \varepsilon^{12} = 1$. We restrict
ourselves to quadratic terms in fermions.

%Covariant derivatives are
%defined by e.g.
% \begin{align}
%  D_\mu \xi^m & = \partial_\mu \xi^ m + \omega_{\mu,}{}^m{}_n \xi^n \,,
%  \notag \\
%  D_\mu \eps^i & = \partial_\mu \eps^i + \tfrac14 \omega_{\mu,mn} \Gamma^{mn}
%\eps^i \,.
% \end{align}

\subsection{The ungauged case} \label{ungauged-section}

The graviton multiplet for minimal five-dimensional supergravity consists of the
 F\"{u}nfbein $e_\mu{}^a$,
 a symplectic Majorana gravitino $\psi_{\mu i}$ and a vector
 $A_\mu$. The dynamics is governed by the Lagrangian
 \begin{align}
  {\mathcal L} = & \sqrt{g} [ - \tfrac12 R - \tfrac14 F_{\mu \nu}
  F^{\mu \nu} - \tfrac12 \bar{\psi}_{\mu}{}^i \Gamma^{\mu \nu \rho}
  D_\nu \psi_{\rho i} - \tfrac{3}{8 \sqrt{6}} i \bar{\psi}_{\mu}{}^i
  ( \Gamma^{\mu \nu \rho \sigma} + 2 g^{\mu \nu} g^{\rho \sigma}  ) F_{\nu \rho} \psi_{\sigma
  i} ] \notag \\
  & + \tfrac{1}{6 \sqrt{6}} \varepsilon^{\mu \nu \rho \sigma
  \lambda} A_\mu F_{\nu \rho} F_{\sigma \lambda} \,,
 \label{action-ung}
 \end{align}
 where the field strength is given by $F_{\mu \nu} = 2 \partial_{[
\mu} A_{\nu]}$ and $D_\mu$ is the covariant derivative with respect
to general coordinate and Lorentz transformations.

The action is invariant under ungauged supersymmetry transformations
 given by
 \begin{align}
  \delta e_\mu{}^m & = \tfrac12 \beps^i \Gamma^m \psi_{\mu i} \,,
  \notag \\
  \delta \psi_{\mu i} & = D_\mu \eps_i + \tfrac{1}{4 \sqrt{6}} i (
  \Gamma_\mu{}^{\nu \rho} - 4 \delta_{\mu}{}^\nu \Gamma^\rho) F_{\nu
  \rho} \eps_i \,, \notag \\
  \delta A_\mu & = - \tfrac{\sqrt{6}}{4} i \beps^i \psi_{\mu i} \,,
 \label{susy-transf-ung}
 \end{align}
The commutator of two supersymmetry transformations generates the
supersymmetry algebra
 \begin{align}
  [\delta_1 , \delta_2] = \delta_{\rm gct} + \delta_{\rm Lorentz}
  + \delta_{\rm susy} + \delta_{\rm gauge}
  + \delta_{\mathcal L} \,,
 \label{susy-alg}
 \end{align}
with the following parameters for the general coordinate, local
Lorentz, supersymmetry and gauge transformations\footnote{We differ
with respect to \cite{gauged} in the sign of the third term
of the Lorentz transformation.}:
 \begin{align}
  \xi^\mu & = \tfrac12 \beps_1^i \Gamma^\mu \eps_{2i} \,, \notag \\
  \Lambda^{mn} & = \xi^\nu \omega_{\nu}{}^{mn} + \tfrac{1}{4
  \sqrt{6}} i
  \beps_1^i (\Gamma^{mnpq} + 4 g^{mp} g^{nq} ) F_{pq}
  \eps_{2i} \,, \notag \\
  \eta^i & = - \xi^\mu \psi_{\mu i} \,, \notag \\
  \lambda^{(0)} & = - \tfrac{\sqrt{6}}{4} i \beps_1^i \eps_{2i} -
  \xi^\nu A_\nu \,.
 \end{align}
Here we use the following conventions:
 \begin{align}
  \delta_{\rm gct} A_\mu & = - \xi^\nu \partial_\nu A_\mu -
  A_\nu \partial_\mu \xi^\nu \,, \notag \\
  \delta_{\rm Lorentz}  e_\mu{}^m & = - \Lambda^m{}_n
  e_\mu{}^n \,, \notag \\
  \delta_{\rm gauge}  A_\mu & = - \partial_\mu
  \lambda^{(0)} \,,
 \end{align}
with the obvious generalisation of general coordinate
transformations to other forms.

The last term in \eqref{susy-alg} is a possible first-order field equation that
 can occur when closing the algebra. This is a common feature for the fermions,
 on which supersymmetry only closes modulo their equations of motion.
 In the following we will also find first-order constraints when realising the
 supersymmetry algebra on tensors of higher rank.

 In addition to the local symmetries discussed above, the theory
 also has a global $SU(2)$ R-symmetry. This symmetry only acts on the
 gravitino (in the fundamental representation) while the metric and vector are invariant under it.

We now would like to see whether one can realise the supersymmetry
algebra on other fields as well. We start with a tensor and make the
following Ans\"{a}tze for the transformation under supersymmetry:
 \begin{align}
  \delta B_{\mu \nu} = b_1 \beps^i \Gamma_{[ \mu} \psi_{\nu ] i}
  + b_2 A_{[ \mu} \delta A_{\nu]} \,.
 \end{align}
One finds that the supersymmetry algebra closes provided $b_1 =
\tfrac34 b_2 = - \tfrac12 \sqrt{6}$ and up to both
the gauge transformations
 \begin{align} \label{transformation}
  \delta_{\rm gauge} B_{\mu \nu} =
  - 2 \partial_{[ \mu} \lambda^{(1)}_{\nu ]} - \tfrac13 \sqrt{6} \lambda^{(0)} F_{\mu \nu}
  \,, \quad
  \lambda^{(1)}_\nu = - B_{\nu \sigma} \xi^\sigma + \tfrac14 \sqrt{6}
   \beps_1^i \Gamma_\nu \eps_{2i} - \tfrac12 i \beps_1^i \eps_{2i} A_\nu \,,
 \end{align}
and the duality relation, or first-order field equation,
 \begin{align}
   \delta_{\mathcal L} B_{\mu \nu} = - (H_{\mu \nu \rho} - \tfrac{1}{2} \sqrt{-g}
   \varepsilon_{\mu \nu \rho
   \sigma \lambda} F^{\sigma \lambda}) \xi^\rho \,, \quad
   H_{\mu \nu \rho} = 3 \partial_{[\mu} B_{\nu \rho]} - \sqrt{6}
   A_{[\mu} F_{\nu \rho]} \,.
 \label{duality-relation}
 \end{align}
Since this has to vanish for all supersymmetry transformations we have
to require the equation in brackets to vanish. Indeed, from this duality
relation follows the field equation for
the vector
 \begin{align}
  \nabla^\mu F_{\mu \nu} = - \frac{1}{2\sqrt{6}} \sqrt{-g}
  \varepsilon_{\nu \mu_1 \ldots \mu_4} F^{\mu_1 \mu_2} F^{\mu_3
  \mu_4} \,,
  \label{field-eq}
 \end{align}
which can also be derived from the action \eqref{action-ung}. Hence
we conclude that it is possible to realise supersymmetry on a
tensor, provided it is the Hodge dual to the vector. Summing up, the
supersymmetry algebra only closes up to the duality relation
\eqref{duality-relation}, which can be seen as a bosonic first-order
field equation.

Turning to higher-rank anti-symmetric tensors, it can be seen that the algebra only allows for supersymmetry transformations of the form $\beps^i \Gamma_{[\mu_i \cdots \mu_n} \psi_{\mu_{n+1}]}^j$ which are anti-symmetric in $i$ and $j$ when $n=0,1$ and symmetric when $n=2,3$ (mod $4$). Therefore we make the following Ans\"{a}tze:
 \begin{align}
  \delta C^{ij}_{\mu \nu \rho} & = i c_1 \beps^{(i} \Gamma_{[\mu \nu}
  \psi_{\rho]}^{j)} \,, \notag \\
  \delta D^{ij}_{\mu \nu \rho \sigma} & = d_1 \beps^{(i} \Gamma_{[ \mu \nu
  \rho} \psi_{\sigma]}^{j)} + d_2 A_{[\mu} \delta C^{ij}_{\nu \rho \sigma]} \,, \notag \\
  \delta E_{\mu \nu \rho \sigma \tau} & = i e_1 \beps^i \Gamma_{[ \mu
  \nu \rho \sigma} \psi_{\tau] i} + e_2 A_{[ \mu} B_{\nu \rho} \delta B_{\sigma \tau
  ]} \,,
 \end{align}
where the first two lines are symmetric in $i$ and $j$.
Note that we could have included more terms, e.g.~$C^{ij} \wedge \delta A$ in $\delta D^{ij}$,
but these can be absorbed into a redefinition of $D^{ij}$. The above Ans\"{a}tze are the
most general modulo such redefinitions. In addition we can impose the symplectic reality conditions
 \begin{align}
  C^{ij} - C_{ij}^* = D^{ij} - D_{ij}^* = 0 \,. \label{symplectic}
 \end{align}
It can be verified that these conditions are invariant under the
above supersymmetry transformations and under the $SU(2)$
R-symmetry. Under the latter these higher-rank tensors therefore
transform as triplets\footnote{In the first preprint version of this paper
we only considered the trace of these symmetric representations. This
is not $SU(2)$ covariant, as was correctly pointed out afterwards in
\cite{Riccioni:2007hm}. However, the introduction of the triplet
representations above does give an $SU(2)$-covariant formulation.}. Note that the original bosonic fields (i.e.~the metric and the vector) are invariant under the $SU(2)$ symmetry; until the introduction of the higher-rank tensors this is a symmetry that only acts on the fermionic sector of the theory.

The closure of the supersymmetry algebra on these higher-rank tensors requires the following constants:
 \begin{align}
  c_1 d_2 = - \sqrt{6} d_1 \,, \quad e_2 = 0 \,,
 \end{align}
and associated gauge transformations with parameters:
 \begin{align}
  \delta_{\rm gauge} C^{ij}_{\mu \nu \rho} & =
  - 3 \partial_{[ \mu} \lambda^{(2)ij}_{\nu \rho ]} \,,
  \quad \lambda^{(2)ij}_{\mu \nu} = - C^{ij}_{\mu \nu \rho} \xi^\rho +
  \tfrac13 i c_1 \beps_1^{(i} \Gamma_{\mu \nu} \eps_2^{j)} \,, \notag \\
  \delta_{\rm gauge} D^{ij}_{\mu \nu \rho \sigma} & =
  - 4 \partial_{[ \mu} \lambda^{(3)ij}_{\nu \rho \sigma]} \,,
   \quad \lambda^{(3)ij}_{\mu \nu \rho}  =
   - D^{ij}_{\mu \nu \rho \sigma} \xi^\sigma - \tfrac14 d_1 (\beps_1^{(i}
   \Gamma_{\mu \nu \rho} \eps_2^{j)} - \sqrt{6} i A_{[\mu} \beps_1^{(i}
   \Gamma_{\nu \rho ]} \eps_2^{j)}) \,, \notag \\
  \delta_{\rm gauge} E_{\mu \nu \rho \sigma \tau} & = - 5 \partial_{[ \mu}
  \lambda^{(4)}_{\nu \rho \sigma \tau]} \,, \quad \lambda^{(4)}_{\mu \nu \rho \sigma}
   = - E_{\mu \nu \rho \sigma \tau} \xi^\tau + \tfrac15 i e_1 \beps_1^i
   \Gamma_{\mu \nu \rho \sigma}  \eps_{2i} \,.
   \label{higher-gauge-transformations}
 \end{align}
In addition, on the right hand side of the supersymmetry algebra
appear the following first-order field equations for the three- and
four-forms:
 \begin{align}
  \delta_{\mathcal L} C^{ij}_{\mu \nu \rho} = - (4 \partial_{[ \mu} C^{ij}_{\nu \rho \sigma]}) \xi^\sigma \,,
  \quad
  \delta_{\mathcal L} D^{ij}_{\mu \nu \rho \sigma} =
  - (5 \partial_{[ \mu} D^{ij}_{\nu \rho \sigma \tau]}) \xi^\tau \,, \label{vanishing-fs}
 \end{align}
which imply that their curvatures vanish, i.e.~these potentials are closed. In combination with their gauge transformations this implies that they do not carry any local degrees of freedom\footnote{A similar phenomenon, gauge vectors with vanishing field strengths and no local degrees of freedom, was encountered in \cite{Samtleben} in the context of $d=2$ supergravity.}. Indeed, they can only be relevant in topologically non-trivial manifolds, e.g.~when they are proportional to volume forms of non-contractible cycles.

A complementary conclusion can be reached for the five-form $E$. Its supersymmetry transformation is proportional to that of the Levi-Civita tensor, which is
 \begin{align}
  \delta ( \sqrt{-g} \varepsilon_{\mu \nu \rho \sigma \tau} ) & = - \tfrac52 i \beps^i \Gamma_{[ \mu
  \nu \rho \sigma} \psi_{\tau] i} \,,
 \end{align}
and hence $E$ is not an independent field but rather composed of the metric, i.e.~it it proportional to the volume form of space-time: $E = - \tfrac25 e_1 \varepsilon$. Indeed, with this identification $\lambda^{(4)}$
vanishes automatically, consistent with the absence of a gauge transformation for
the Levi-Civita tensor.

Hence there are no local degrees of freedom associated to the potentials $C^{ij}$ and $D^{ij}$ and
there is no independent five-form potential $E$. It is interesting to note that the commutator
of two susy transformations on these
potentials turns out to be given by a gauge transformation:
 \begin{align}\label{totalderivative}
  [\delta_1 , \delta_2 ] C^{ij}_{\mu \nu \rho} & =
  - 3 \partial_{[ \mu} \tlambda^{(2)ij}_{\nu \rho ]} \,,
  \quad \tlambda^{(2)ij}_{\mu \nu}  =
  \tfrac13 i c_1 \beps_1^{(i} \Gamma_{\mu \nu} \eps_2^{j)} \,, \notag \\
  [\delta_1 , \delta_2 ] D^{ij}_{\mu \nu \rho \sigma} & =
  - 4 \partial_{[ \mu} \tlambda^{(3)ij}_{\nu \rho \sigma]} \,,
   \quad \tlambda^{(3)ij}_{\mu \nu \rho}  = - \tfrac14 d_1 (\beps_1^{(i}
   \Gamma_{\mu \nu \rho} \eps_2^{j)} - \sqrt{6} i A_{[\mu} \beps_1^{(i}
   \Gamma_{\nu \rho ]} \eps_2^{j)} ) \,, \notag \\
  [\delta_1 , \delta_2 ] E_{\mu \nu \rho \sigma \tau} & = - 5 \partial_{[ \mu}
  \tlambda^{(4)}_{\nu \rho \sigma \tau]} \,, \quad \tlambda^{(4)}_{\mu \nu \rho \sigma}
   = \tfrac15 i e_1 \beps_1^i
   \Gamma_{\mu \nu \rho \sigma}  \eps_{2i} \,.
 \end{align}
One finds that the commutator of supersymmetry does not lead to any terms involving the parameter
$\xi^\mu$ of general coordinate transformations. These terms cancel separately on the right hand
side of the supersymmetry algebra \eqref{susy-alg} due to the contribution \eqref{vanishing-fs}.
Hence the supersymmetry algebra \eqref{susy-alg} is realised in a rather trivial way on these
potentials. Indeed, due to the above commutators, setting $C^{ij}$ and $D^{ij}$ to zero by the gauge transformations \eqref{higher-gauge-transformations} is consistent with supersymmetry.

The presence of the triplets of three- and four-forms, on which
supersymmetry can be realised provided they have vanishing
curvature, may have come as a surprise at this point. In the next
subsection we will see however that they are necessary for the
inclusion of a gauge coupling constant.

\subsection{The gauged case}

We now consider the gauging of a $U(1)$ subgroup of the $SU(2)$
R-symmetry group, with coupling constant\footnote{Here we have
chosen a specific embedding of the gauged $U(1)$ in $SU(2)$ without
loss of generality. To describe the other embeddings one should
replace $g \delta_{ij}$ by $g_{ij}$, which is symmetric and subject
to a symplectic reality condition like \eqref{symplectic}.} $g$
\cite{gauged}. The action for this gauged supergravity is
  \begin{align}
  {\mathcal L} = & \sqrt{g} [ - \tfrac12 R - \tfrac14 F_{\mu \nu}
  F^{\mu \nu} - \tfrac12 \bar{\psi}_{\mu}{}^i \Gamma^{\mu \nu \rho}
  (D_\nu \psi_{\rho i} - g A_\nu \delta_{ij} \psi_\rho^j) + \notag \\
  & \;\;\;\;\;\; - \tfrac{3}{8 \sqrt{6}} i \bar{\psi}_{\mu}{}^i
  ( \Gamma^{\mu \nu \rho \sigma} + 2 g^{\mu \nu} g^{\rho \sigma}  ) F_{\nu \rho} \psi_{\sigma
  i} - \tfrac14 \sqrt{6} i g \bar{\psi}_\mu^i \Gamma^{\mu \nu} \psi_\nu^j \delta_{ij} + 4 g^2] \notag \\
  & + \tfrac{1}{6 \sqrt{6}} \varepsilon^{\mu \nu \rho \sigma
  \lambda} A_\mu F_{\nu \rho} F_{\sigma \lambda} \,,
 \label{action-g}
 \end{align}
where the field strength is still given by $F_{\mu \nu} = 2 \partial_{[
\mu} A_{\nu]}$. These are invariant under the following supersymmetry variations:
 \begin{align}
  \delta e_\mu{}^m & = \tfrac12 \beps^i \Gamma^m \psi_{\mu i} \,,
  \notag \\
  \delta \psi_{\mu i} & = D_\mu \eps_i + \tfrac{1}{4 \sqrt{6}} i (
  \Gamma_\mu{}^{\nu \rho} - 4 \delta_{\mu}{}^\nu \Gamma^\rho) F_{\nu
  \rho} \eps_i - g A_\mu \delta_{ij} \eps^j - \tfrac{1}{\sqrt{6}} i g \Gamma_\mu \delta_{ij} \eps^j \,, \notag \\
  \delta A_\mu & = - \tfrac{\sqrt{6}}{4} i \beps^i \psi_{\mu i} \,.
 \label{susy-transf-g}
 \end{align}
Note that there are only corrections to the supersymmetry variation of the
fermion and not to those of the metric and vector.

It turns out that the supersymmetry variations of all higher-rank potentials are unchanged as well, i.e.~equal to their ungauged expressions, just like the other bosons in \eqref{susy-transf-g}. The only differences appear on the right hand side of the supersymmetry algebra for the potentials $B$ and $D^{ij}$: the two-form gauge transformation becomes
 \begin{align}
  \delta_{\rm gauge} B_{\mu \nu} & = - 2 \partial_{[ \mu} \lambda^{(1)}_{\nu ]} - \tfrac13 \sqrt{6} \lambda^{(0)} F_{\mu \nu} +
   \beta g \lambda^{(2)ij}_{\mu \nu} \delta_{ij}
  \,, \label{gauged-transformation}
 \end{align}
where $\beta = \sqrt{6} b_1 / c_1$ and the duality relations (or first-order field equations) for $B$
and $D^{ij}$ become
 \begin{align}
   \delta_{\mathcal L} B_{\mu \nu} & = - (H_{\mu \nu \rho} - \beta g C^{ij}_{\mu \nu \rho} \delta_{ij} - \tfrac{1}{2} \sqrt{-g} \varepsilon_{\mu \nu \rho
   \sigma \lambda} F^{\sigma \lambda}) \xi^\rho \,, \notag \\
  \delta_{\mathcal L} D^{ij}_{\mu \nu \rho \sigma} & = - (5 \partial_{[ \mu} D^{ij}_{\nu \rho \sigma \tau]} - \tfrac12 \gamma g \delta^{ij} E_{\mu \nu \rho \sigma \tau} ) \xi^\tau \,,
 \end{align}
where $\gamma = - \tfrac53 \sqrt{6} d_1 / e_1$.

Note that the trace part of the field strength of the four-forms
potential $D^{ij}$ is non-vanishing  in the gauged theory. This
implies that this potential, unlike in the ungauged case, can no
longer be gauged away locally. Recalling the identification of $E$
with the Levi-Civita tensor, the duality relation for the trace of
the four-form $D^{ij}$ implies that its field strength is Hodge dual
to the mass parameter or gauge coupling
 constant $g$. This is analogous to the identification of e.g.~the
 field strength of the nine-form in IIA supergravity \cite{nineform1, nineform2}
 with Romans' mass parameter \cite{Romans}. Hence the presence of the
 four-forms
 in the supersymmetry algebra is directly related to the possibility
 of gauging the $U(1)$ group. This explains why $D^{ij}$ also appeared in
  the ungauged case. Indeed, its appearance there can be seen as
  a necessary condition for and hence a prediction of the existence of gauged supergravity.

In the same spirit, the gauging explains the presence of the
three-forms $C^{ij}$ in the superalgebra. Their gauge
transformations are  necessary to be able to realise supersymmetry
on the tensor in the gauged case, since the latter transforms under
the former. Indeed, the tensor $B$ is pure gauge due to the
$\lambda^{(2)ij}$ term in its gauge transformation. When gauging
away $B$, the associated degrees of freedom are carried by the trace
of $C^{ij}$. It has a vanishing field strength but its gauge freedom
has been fixed, giving rise to the same number of local degrees of
freedom as a two-form gauge potential. Alternatively, we could
locally choose to set $C^{ij}$ to zero, but in order to preserve
this gauge choice under the commutator of supersymmetry we need a
compensating transformation $\tlambda^{(2)ij}$ given by
\eqref{totalderivative}. Also note that, although the field strength
$H$ contains a term $g C^{ij} \delta_{ij}$, this does does not
modify the field equation \eqref{field-eq}, in accordance with the
above action for the gauged case.

Even though $C^{ij}$ still has vanishing curvature and the
commutator of supersymmetry acts as a total derivative on it, the
three-forms turns out to play a crucial role in dualising the vector
into a tensor when $g \neq 0$. Indeed, it is impossible to realise
supersymmetry on the tensor without including $C^{ij}$.
 This is in contrast to the ungauged case, where it is consistent to
 consider only potentials up to a certain rank. In the gauged case such a
 hierarchy is no longer present: a higher-rank potential can be necessary to
  realise supersymmetry on a potential of lower rank, as we have found for $B$ and $C^{ij}$.

Summarising, we have found in this subsection that the presence of
$C^{ij}$ and $D^{ij}$ in the supersymmetry algebra are both related
to the gauging: the four-forms predict the possibility to include a
gauging, while the three-forms are necessary to dualise the vector
in the gauged case. The latter seems to be a novel mechanism that we
have not encountered in the literature\footnote{For instance, in the
formalism of \cite{deWit} for gauged $d=5$ maximal supergravities,
dual tensors are also introduced and the supersymmetry algebra only
closes up to first-order duality relations for them, but there are
no terms like $g \lambda^{(2)}$ in their gauge transformations.}.

\section{Very-extended $G_2$}

In this section we will recapitulate the predictions from very
extended $G_2$ and compare against the findings from the
supersymmetry algebra.

Given a Kac-Moody algebra which is the very extension of some Lie
algebra, one can decompose its adjoint representation into
representations of a Lie subalgebra $A_n$ (the 'gravity line').
These are labelled by their level $l$ in the Kac-Moody algebra and
can be interpreted to correspond with fields in $d = n + 1$
dimensions. This has been explained in e.g.~\cite{very extended} and references therein, where more details can be found.

\begin{figure}[ht]
\begin{center}
\includegraphics[height=1.5cm]{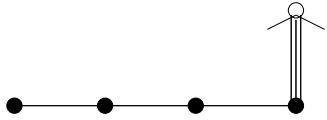}
\caption{\label{g2pppdynk}The extended Dynkin diagram of $G_2^{+++}$
with its horizontal $A_4$ subalgebra.}
\end{center}
\end{figure}

The relevant very extended algebra in the present case is $G_2^{+++}$, whose extended Dynkin diagram is given in the figure. Its decomposition in $A_4$ representations has been given in \cite{very extended}, from which
we copy the relevant table. Note that there is no internal $SU(2)$ symmetry in addition to the space-time $A_4$ symmetry, and hence all ensuing representations will be singlets of $SU(2)$.

\begin{table}[ht]
\begin{center}
\begin{tabular}{cccrrrr}
$l$&$A_4$ weight&$G_2^{+++}$ element
$\alpha$&$\alpha^2$&$ht(\alpha)$&$\mu$ & Interpretation \\
\hline
0&[1,0,0,1]&(1,1,1,1,0)&2&4&1 & graviton \\
1&[0,0,0,1]&(0,0,0,0,1)&2&1&1 & vector $A$ \\
2&[0,0,1,0]&(0,0,0,1,2)&2&3&1 & tensor $B$ \\
3&[0,0,1,1]&(0,0,0,1,3)&6&4&1 & dual graviton \\
3&[0,1,0,0]&(0,0,1,2,3)&0&6&0 & \\
4&[0,1,0,1]&(0,0,1,2,4)&2&7&1 & mixed \\
4&[1,0,0,0]&(0,1,2,3,4)&-4&10&0 & \\
5&[0,1,1,0]&(0,0,1,3,5)&2&9&1 & mixed\\
5&[1,0,0,1]&(0,1,2,3,5)&-4&11&1 & mixed\\
5&[0,0,0,0]&(1,2,3,4,5)&-10&15&0 & \\
\end{tabular}
\end{center} \caption{\label{g2pppdec} The first levels of the decomposition of
$G_2^{+++}$ with respect to $A_4$. All representations are $SU(2)$ singlets.}
\end{table}

The space-time field interpretation for the first four entries is as
graviton, vector and tensor, respectively. These agree with our
results in the previous section, where we found that one can realise
supersymmetry on $e_\mu{}^a$, $A_\mu$ and $B_{\mu \nu}$. The fourth should correspond to the dual graviton, which we did not consider since it has mixed symmetries and we restrict ourselves to anti-symmetric tensors. The
remaining entries  either have mixed symmetries or are absent (with
vanishing multiplicity $\mu$). At higher levels $l \geq 6$ there are
only representations with more than six space-time indices.

Note in particular that there are no four-form
potentials\footnote{The same absence was noted by \cite{Mizoguchi2}
in the context of a one-dimensional $\sigma$-model based on
overextended $G_2$. There it was interpreted as predicting the
absence of $R^2$ higher-order corrections, which however do occur
for this supergravity. This paradox may be resolved by the
observation of \cite{weights1, weights2} that higher-order
corrections correspond to weights instead of roots of the
overextended algebra.} predicted by very extended $G_2$. This is in
clear contradistinction to the results from the supersymmetry
algebra, which does allow for a triplet of four-forms whose field
strength is dual to the gauge coupling constant. Hence it emerges
that very extended $G_2$ should be associated to ungauged $d=5$
minimal supergravity and not to the corresponding gauged
supergravity. In addition to the absence of the four-forms, there is
also no five-form predicted by very extended $G_2$. This agrees with
both the ungauged and the gauged supersymmetry algebra.

Given that $G_2^{+++}$ is associated to the ungauged case, the
vector can be identified as a raising operator from which the entire
bosonic gauge algebra of the ungauged theory can be generated. To
see this one must first make the following redefinition of the gauge
algebra. As things stand, the gauge transformation
\eqref{transformation} is Abelian and non-local, due to the term
proportional to $F$. One can redefine the gauge parameter by
${\lambda}^{(1)'} = \lambda^{(1)} + \tfrac13 \sqrt{6} \lambda^{(0)}
A$ to obtain the transformation
  \begin{align} \label{nA-transformation}
  \delta_{\rm gauge} B_{\mu \nu} =
  - 2 \partial_{[ \mu} {\lambda}^{(1)'}_{\nu ]} +
  \tfrac23 \sqrt{6} \partial_{[\mu} \lambda^{(0)} A_{\nu ]}
  \,,
  \end{align}
which is non-Abelian and local. A similar phenomenon was observed in \cite{IIA-superalgebra},
 where the non-Abelian gauge algebra was interpreted in terms
 of raising operators. In our case these are the gauge transformation
 $\bf 1$ of the vector, and we sketchily have
\begin{align}
  \left [{\bf 1}, {\bf 1} \right ] = {\bf 2} \,,
\end{align}
where $\bf 2$ is the gauge transformation of the tensor. Hence the
vector can be interpreted as the raising operator $\bf 1$, in
agreement with the fact that the node outside of the gravity line is
at the outer right position in the extended Dynkin diagram, and all
other gauge transformations can be generated by considering multiple
commutators of it. For instance, the double commutator $[ [{\bf 1},
{\bf 1}] , {\bf 1}]$ should give rise to the gauge transformation of
the dual graviton, see also \cite{Bergshoeff}. From this point of
view it also follows that the multiple commutators of the singlet
$\bf 1$ can not give rise to the gauge transformations of the
triplets of higher-rank forms.

\section{Discussion} \label{discussion}

In this note we have compared the possibilities to realise the
$N=2$, $d=5$ supersymmetry algebra on higher-rank tensors with the
predictions of very extended $G_2$. Our main results are the
inclusion of triplets of three- and four-forms in the supersymmetry
algebra, necessary for the gauging of the $U(1)$, and the failure of
very extended $G_2$ to capture these forms.

The absence of the four-forms in very extended $G_2$ is in contrast
to the previously considered case of $E_{11}$ and gaugings of
maximal supergravities, where the very extended algebra contains
$(d-1)$-forms corresponding to the possible gauge coupling constants
or mass parameters. 
A caveat here is that there are more deformations allowed for by supersymmetry which are not captured by
$E_{11}$, that correspond to the gauging of the 'trombone' or scale
symmetry of the field equations and Bianchi identities \cite{HLW,
trombone9D, trombone8D}. These are not symmetries of the Lagrangian,
and indeed their gauging leads to field equations that cannot be
derived from an action principle. In addition, these symmetries are
expected to be broken by higher-order corrections. The situation
considered here is therefore of a different nature: gauging the
$U(1)$ leads to a perfectly bonafide gauged supergravity with an
action principle. 
It does differ from gauged maximal supergravity in that
its original bosonic fields are invariant under the symmetry that is gauged, while the gauge groups of maximal supergravity do act on the original bosonic sector \cite{Riccioni:2007hm}.

The absence of the $U(1)$ gauging is all the more striking from the following point of view.
The $N=2$ gauged supergravity can be obtained as a truncation of $N=8$ supergravity
with an $SO(6)$ gauging \cite{Gunaydin}, which is included in $E_{11}$.
The gauge coupling constant survives the truncation from $N=8$ to minimal $N=2$ pure supergravity.
From the very extended algebras point of view, $E_{11}$ can be truncated to very extended $G_2$.
This works flawlessly for the propagating degrees of freedom, but the gauge coupling constant
is lost in the process. This suggests that there is a different truncation of $E_{11}$, which contains both $G_2^{+++}$ and the $SU(2)$ triplets of three- and four-forms generators, and therefore accounts for both propagating and non-propagating degrees of freedom. It would be interesting to uncover whether such an algebra exists and what its structure is. 

In this note we have presented an example with eight supercharges and the
non-simply laced $G_2^{+++}$, where the very extended algebra does not capture the possible gauging of the supergravity theory. Note that this is even without including any matter multiplets, which is an additional option in less than maximal supergravity. It will be very interesting to extend this analysis to other cases, with other supergravities and very extended algebras, and to investigate what the requirements or reasons are for the
non-propagating degrees of freedom to be present or absent in the very extended algebras. In the latter case, one could also look for possible extensions of these algebras that do contain all non-propagating degrees of freedom, similar to a possible truncation of $E_{11}$ that extends $G_2^{+++}$ with the triplets of generators.

We have also observed that in the gauged case the supersymmetry
algebra does not preserve the level structure. That is, the
commutator of supersymmetry on a form can receive gauge
contributions from a higher-rank form, in our case $B$ and $C^{ij}$.
For this reason it is not always possible to only include fields up
to a certain level $l$. One may expect this to be a general
phenomena that will also occur for level decompositions in other
theories.

\section*{Acknowledgements}

We are grateful to Eric Bergshoeff, Teake Nutma and Fabio Riccioni for very stimulating
discussions. This work has been supported by the European EC-RTN project
MRTN-CT-2004-005104, MCYT FPA 2004-04582-C02-01 and CIRIT GC
2005SGR-00564. We thank the Galileo Galilei Institute for
Theoretical Physics for its hospitality and INFN for partial support
during the completion of this work.

\end{document}